\documentclass[aps,prb,twocolumn,showpacs,superscriptaddress,floatfix]{revtex4}

\pdfoutput = 1

\usepackage{graphicx}
\usepackage{dcolumn}
\usepackage{bm}
\usepackage{stmaryrd}
\usepackage{latexsym}
\usepackage{amssymb}
\usepackage{amsfonts}
\usepackage{amsmath}
\usepackage{dsfont}
\usepackage{ulem}

\begin{document}
\title{Compressibility of bilayer graphene}

\author{Giovanni Borghi}
\affiliation{International School for Advanced Studies (SISSA), via Beirut 2-4, I-34014 Trieste, Italy}
\author{Marco Polini}
\email{m.polini@sns.it} \homepage{http://qti.sns.it}
\affiliation{NEST, Istituto Nanoscienze-CNR and Scuola Normale Superiore, I-56126 Pisa, Italy}
\author{Reza Asgari}
\affiliation{School of Physics, Institute for Research in Fundamental Sciences (IPM), Tehran 19395-5531, Iran}
\author{A.H. MacDonald}
\affiliation{Department of Physics, The University of Texas at Austin, Austin, Texas 78712, USA}

\begin{abstract}
Bilayer graphene is a recently isolated and intriguing class of many-body systems with massive chiral quasiparticles. We present theoretical results  
for the electronic compressibility of bilayer graphene that are based on a
four-band continuum band structure model combined with a random phase approximation treatment of electronic correlations. 
We find that the compressibility is strongly suppressed by electron-electron interactions at low carrier densities.
Correlations do not lead to any qualitative new features, but are crucially important for a quantitative understanding 
of this fundamental thermodynamic property of graphene bilayers. 
\end{abstract}


\maketitle

\section{Introduction} 
\label{sect:intro}

Crystalline bilayers of graphene (BLG) produced by mechanical exfoliation of thin graphite or by 
thermal decomposition of silicon carbide have recently attracted a great deal of attention because of their 
many unique electronic properties~\cite{reviews,ohta_science_2006,novoselov_naturephys_2006,castro_prl_2007,rutter_science_2007,oostinga_naturemat_2008,
feldman_natphys_2009,zhao_prl_2010}. 
BLG quasiparticles behave at low energies like massive chiral fermions~\cite{mccann_prl_2006} and exhibit a plethora of interesting 
properties, including broken-symmetry states at very weak magnetic fields when the bilayer is suspended~\cite{feldman_natphys_2009} 
to reduce disorder, and anomalous exciton condensation in the quantum Hall regime~\cite{barlas_prl_2010}.

Since BLG consists of two single-layer graphene (SLG) systems separated
by a small distance $d \sim 3.35$~\AA, one expects inter-layer electron-electron interactions 
be crucial to the physics of this system. With this motivation, many-body effects in BLG have already been 
studied by several authors~\cite{nilsson_prb_2006,wang_prb_2007,kusminskiy_prl_2008,kusminskiy_EPL_2009,hwang_prl_2008,borghi_ssc_2009,toke_condmat_2009,borghi_prb_2009,barlas_prb_2009,wang_prb_2010}. Particular attention has been devoted to the study of interaction effects close to charge neutrality~\cite{min_prb_2008,zhang_prb_2010,vafek_prb_2010,nandkishore_prl_2010,guinea_physics_2010} (see also Ref.~\onlinecite{sun_prl_2009}) where it has been shown that BLG is prone to a number of interesting instabilities, including sublattice pseudospin ferromagnetism - a type of orbital order which leads to spontaneous inversion symmetry~\cite{min_prb_2008,zhang_prb_2010} breaking.

Thermodynamic quantities such as the electronic compressibility $\kappa$ or the spin susceptibility $\chi_{\rm S}$ are very powerful probes of exchange and correlation effects in interacting many-electron systems~\cite{Giuliani_and_Vignale} since they are intimately linked with the equation of state. The electronic compressibility of a conventional parabolic-band two-dimensional (2D) electron gas was first measured by Eisenstein {\it et al.}~\cite{eisenstein_prl_1992} in 1992. For sufficiently low densities, and zero magnetic fields, it was found that the inverse thermodynamic density-of-states, which is proportional to $1/\kappa$, changes sign becoming {\it negative}, a fact that can be easily explained by properly including exchange contributions to the free-electron equation of state~\cite{Giuliani_and_Vignale}.  In an ordinary 2D electron gas corrections to the compressibility due to correlation effects omitted in Hartree-Fock approximations are relatively small.  
The ``field penetration technique" introduced in Ref.~\onlinecite{eisenstein_prl_1992} and later discussed in great detail in Ref.~\onlinecite{eisenstein_prb_1994} actually uses a double-layer 2D electron system made up of {\it two} closely-spaced 2D electron gases 
and can also be used to accurately measure the compressibility of BLG.

In this work we present a calculation of the electronic compressibility of BLG based on the four-band continuum model. We include beyond-Hartree-Fock correlation contributions to the ground-state energy by using a random phase approximation. We demonstrate that the correlation 
contribution to the compressibility in BLG is crucial when dielectric screening is weak and interactions within the graphene sheet are 
strong.  Indeed, neglect of correlation effects leads to an error of the order of $100$\% in the case of suspended bilayers.
We compare our results for the compressibility of BLG with those obtained earlier for SLG~\cite{barlas_prl_2007} and are able to
clearly identify the physical origin of the main differences. For simplicity we assume here that the bilayer remains in a normal Fermi-liquid state down to very low densities. The behavior of the compressibility when one of the exotic low-density phases predicted in Refs.~\onlinecite{min_prb_2008,zhang_prb_2010,vafek_prb_2010,nandkishore_prl_2010,guinea_physics_2010} is approached from the high-density Fermi-liquid phase is beyond the scope of the present theory.

We note that the compressibility of BLG has already been calculated at the Hartree-Fock (HF) level in Ref.~\onlinecite{kusminskiy_prl_2008}. We comment on the relationship between our results and those obtained in this earlier work in Sect.~\ref{sect:mainresults}.  We restrict our attention in this article to the case of a balanced bilayer in which inversion symmetry is not broken by an electrical potential difference between the layers.  When a potential difference is present a gap opens up in the single-particle energy spectrum~\cite{mccann_prl_2006}; there is no gap between conduction and valence bands in the balanced bilayer limit that we consider. We also neglect trigonal warping effects in the bands which become important only at very low densities at which disorder effects normally dominate. Both limitations are shared with the HF theory of Ref.~\onlinecite{kusminskiy_prl_2008}.

Our paper is organized as follows. In Sect.~\ref{sect:modelH} we introduce the model and the linear-response functions which control 
BLG ground-state properties. In Sect.~\ref{sect:kappaxc} we (i) derive explicit expressions for the exchange and correlation energies using the integration-over-coupling-constant algorithm and the fluctuation-dissipation theorem, (ii) introduce the random phase approximation for the correlation energy, and (iii) present and comment on our main numerical results. Finally, in Sect.~\ref{sect:conclusions} we summarize our main findings and discuss their signifigance.  Some technical details are relegated to an appendix.

\section{Model Hamiltonian and linear-response functions} 
\label{sect:modelH}

BLG is modeled as two SLG systems separated
by a distance $d$ and coupled by both inter-layer hopping and Coulomb interactions.  Most of the properties we discuss below
depend qualitatively on the Bernal stacking arrangement 
in which one sublattice (say $A$) of the top layer is a near-neighbor of the opposite 
sublattice (say $B$) of the bottom layer. Neglecting trigonal warping, the continuum model single-particle Hamiltonian~\cite{mccann_prl_2006, borghi_prb_2009} of a single valley is ($\hbar=1$),
\begin{equation}\label{eq:kinenergy}
{\hat {\cal T}}= \sum_{{\bm k}, \alpha, \beta} {\hat c}^\dagger_{{\bm k}, \alpha} {\cal T}_{\alpha \beta}({\bm k}) 
{\hat c}_{{\bm k}, \beta}~,
\end{equation}
where ${\cal T}_{\alpha \beta}({\bm k}) $ are the coefficients of the following $4\times 4$ matrix
\begin{equation}\label{eq:kinmatrix}
{\cal T}({\bm k})= -v \gamma^5\gamma^0{\bm \gamma} \cdot {\bm k} - \frac{t_\perp}{2} (\gamma^5\gamma^x +i\gamma^y)~.
\end{equation}
Here $v$ ($\sim 10^{6}~{\rm m}/{\rm s}$) is the Fermi velocity of an isolated graphene layer, 
$t_\perp$ ($\sim 0.35~{\rm eV}$) is the inter-layer hopping amplitude, and 
the $\gamma^\mu$ are $4\times 4$ Dirac $\gamma$ matrices in the chiral representation~\cite{maggiore_book} ($\gamma^5 \equiv -i\gamma^0\gamma^1\gamma^2\gamma^3$). 
The Greek indices $\alpha,\beta$ account for the sublattice degrees of freedom 
in top ($1=A$, $2=B$) and bottom ($3=A$, $4=B$) layers. In the other valley the kinetic Hamiltonian is given by 
${\cal T}'({\bm k}) = {\cal T}^*({-{\bm k}})$. 

If BLG is embedded in a medium with uniform dielectric constant $\epsilon$, 
electrons in the same layer interact {\it via} the 2D Coulomb potential $V_{\rm S}(q)=2\pi e^2/\epsilon q$,
while electrons in different layers interact {\it via} $V_{\rm D}(q)=V_{\rm S}(q)\exp{(-qd)}$. 
If the dielectric media above, below, and between the two layers are not identical these simple expressions are no
longer valid~\cite{profumo_arXiv_2010}.

For practical calculations of thermodynamic quantities and linear-response functions it is convenient to 
work in the single-particle Hamiltonian eigenstate basis.
Diagonalization of ${\cal T}({\bm k})$ yields four hyperbolic bands~\cite{nilsson_prb_2006} with dispersions,
$\varepsilon_{1,2}(k) = \pm \sqrt{v^2 k^2 +t^2_\perp/4} + t_\perp/2$ and 
$\varepsilon_{3,4}(k)  = \pm \sqrt{v^2 k^2 +t^2_\perp/4} - t_\perp/2$.  The interaction contribution to the Hamiltonian is 
\begin{equation}\label{eq:intenergy}
{\hat {\cal H}}_{\rm int} = \frac{1}{2S}\sum_{\bm q} \left[V_+(q){\hat \rho}_{\bm q}{\hat \rho}_{-\bm q} 
+ V_-(q) {\hat \Upsilon}_{\bm q}{\hat \Upsilon}_{-\bm q}\right]~,
\end{equation}
where $S$ is the 2D electron system area,  
$V_\pm = (V_{\rm S} \pm V_{\rm D})/2$, and ${\hat \rho}_{\bm q}$ 
and ${\hat \Upsilon}_{\bm q}$ are respectively the operators for the sum and difference of 
the individual layer densities:
\begin{eqnarray}\label{eq:densop}
{\hat \rho}_{\bm q} &=&\sum_{{\bm k}, \lambda, \lambda'} {\hat c}^{\dagger}_{{\bm k}-{\bm q},\lambda}
({\cal U}^\dagger_{{\bm k}-{\bm q}}{\cal U}_{\bm k})_{\lambda \lambda'}{\hat c}_{{\bm k},\lambda'}
\end{eqnarray}
and 
\begin{eqnarray}\label{eq:upsilonop}
{\hat \Upsilon}_{\bm q} &=&\sum_{{\bm k}, \lambda, \lambda'} {\hat c}^{\dagger}_{{\bm k}-{\bm q},\lambda}
({\cal U}^\dagger_{{\bm k}-{\bm q}} \gamma^5 {\cal U}_{\bm k})_{\lambda \lambda'}{\hat c}_{{\bm k},\lambda'}~.
\end{eqnarray}
Here ${\cal U}_{\bm k}$ is the unitary
transformation from sublattice to band labels $\lambda,\lambda'$ (see Appendix~\ref{app:unitaryU}). 

We evaluate interaction energies using a coupling-constant-integration
scheme which expresses energies in terms of electronic equal-time correlation functions.
The correlation functions can then be related to response functions using 
the fluctuation-dissipation theorem.  When this commonly used approach~\cite{Giuliani_and_Vignale} is adapted to the case of 
BLG, we see from ${\hat {\cal H}}_{\rm int}$ that two response functions are necessary for the evaluation of ground-state properties of BLG~\cite{notationfootnote}: the total-density response function,
\begin{equation}\label{eq:rhorho}
\chi_+(q,\omega) = \frac{1}{S}\langle\langle {\hat \rho}_{\bm q}; {\hat \rho}_{-{\bm q}}\rangle\rangle_\omega~,
\end{equation}
and the density-difference response function
\begin{equation}\label{eq:upsilonupsilon}
\chi_-(q,\omega) = \frac{1}{S}\langle\langle {\hat \Upsilon}_{\bm q}; {\hat \Upsilon}_{-{\bm q}}\rangle\rangle_\omega~.
\end{equation}
Here $\langle\langle {\hat A}; {\hat B} \rangle\rangle_\omega$ is the Kubo product~\cite{Giuliani_and_Vignale},
\begin{equation}
\langle\langle {\hat A}; {\hat B} \rangle\rangle_\omega \equiv -i \lim_{\eta \to 0^+}
\int_0^{+\infty}dt~e^{i \omega t}e^{-\eta t} \langle [{\hat A}(t), {\hat B}]\rangle~,
\end{equation}
$\langle ... \rangle$ being the ground-state expectation value. 

At this point the reader might wonder why we have not introduced the mixed sum and difference response functions, $\langle\langle {\hat \rho}_{\bm q}; {\hat \Upsilon}_{-{\bm q}}\rangle\rangle_\omega/S$ and $\langle\langle {\hat \Upsilon}_{\bm q}; {\hat \rho}_{-{\bm q}}\rangle\rangle_\omega/S$. As explained in Ref.~\onlinecite{borghi_prb_2009}, these response functions vanish because the system Hamiltonian is invariant under spatial inversion (parity).  This is easily seen 
in the sublattice and layer basis where the parity operator ${\cal P}$ is given by 
${\cal P} = (\gamma^x \gamma^5 )^{*}$, with $*$ indicating complex conjugation. 
Using this compact expression for the parity operator ${\cal P}$, we can conveniently calculate its effect on one-body operators, like 
${\hat a}_{\bm q}  = \sum_{{\bm k}, \alpha, \beta}{\hat c}^\dagger_{{\bm k}-{\bm q}, \alpha} {\cal A}_{\alpha \beta}({\bm k}, {\bm q}) 
{\hat c}_{{\bm k}, \beta}$, for example, in the following manner: ${\cal P} {\hat a}_{\bm q} {\cal P} = \sum_{{\bm k}, \alpha, \beta} {\hat c}^\dagger_{{\bm k}-{\bm q}, \alpha} [\gamma^x\gamma^5 {\cal A}({\bm k}, {\bm q}) \gamma^x\gamma^5]^*_{\alpha
\beta}{\hat c}_{{\bm k}, \beta}$. In this way, it is easy confirm that the density-sum operator is even under parity, 
${\cal P} {\hat \rho}_{\bm q} {\cal P} = {\hat \rho}_{\bm q}$, while the density-difference operator is odd, 
${\cal P} {\hat \Upsilon}_{\bm q} {\cal P} = -{\hat \Upsilon}_{\bm q}$. 

Consider now a mixed response function such as $\langle\langle {\hat \rho}_{\bm q}; {\hat \Upsilon}_{-{\bm q}}\rangle\rangle_\omega$. 
We can write it in the exact-eigenstate (Lehmann) representation~\cite{Giuliani_and_Vignale} as
\begin{eqnarray}\label{eq:lehman}
\langle\langle {\hat \rho}_{\bm q}; {\hat \Upsilon}_{-{\bm q}}\rangle\rangle_\omega &=& \sum_{m,n} 
\frac{P_m - P_n}{\omega - \omega_{nm} +i \eta}~\langle \Psi_m|{\hat \rho}_{\bm q}|\Psi_n\rangle \nonumber \\
&\times& \langle \Psi_n|{\hat \Upsilon}_{-{\bm q}}|\Psi_m\rangle~.
\end{eqnarray}
Here $\omega_{nm} =  E_n - E_m$ are excitation energies, $P_n = \exp(-\beta E_n)/{\cal Z}$ [with $\beta= (k_{\rm B} T)^{-1}$ and ${\cal Z}$ the canonical partition function] are Boltzmann factors, $O_{nm} \equiv \langle \Psi_n|{\hat O}|\Psi_m\rangle$ are matrix elements of the operator ${\hat O}$, and the limit $\eta \to 0^+$ is understood. 
We now use that the exact eigenstates $|\Psi_n\rangle$ of the system Hamiltonian are also eigenstates of the parity operator since the the Hamiltonian is parity invariant: ${\cal P}|\Psi_n\rangle = \pm |\Psi_n\rangle$. We find that
\begin{eqnarray}
\langle\langle {\hat \rho}_{\bm q}; {\hat \Upsilon}_{-{\bm q}}\rangle\rangle_\omega &=&\sum_{m,n} 
\frac{P_m - P_n}{\omega - \omega_{nm} +i \eta}~\langle \Psi_m|{\cal P}{\hat \rho}_{\bm q}{\cal P}|\Psi_n\rangle \nonumber \\
&\times& \langle \Psi_n|{\cal P}{\hat \Upsilon}_{-{\bm q}}{\cal P}|\Psi_m\rangle \nonumber \\
&=&-\sum_{m,n} 
\frac{P_m - P_n}{\omega - \omega_{nm} +i \eta}~\langle \Psi_m|{\hat \rho}_{\bm q}|\Psi_n\rangle \nonumber \\
&\times& \langle \Psi_n|{\hat \Upsilon}_{-{\bm q}}|\Psi_m\rangle \nonumber \\
 & =& - \langle\langle {\hat \rho}_{\bm q}; {\hat \Upsilon}_{-{\bm q}}\rangle\rangle_\omega~,
\end{eqnarray}
where we have used that ${\hat \rho}_{\bm q}$ is even under parity while ${\hat \Upsilon}_{\bm q}$ is odd. 
It follows that $\langle\langle {\hat \rho}_{\bm q}; {\hat \Upsilon}_{-{\bm q}}\rangle\rangle_\omega=0$.

In the next Section we will use the two response functions $\chi_+(q,\omega)$ and $\chi_-(q,\omega)$ to calculate exchange and correlation contributions to the BLG equation of state, and thus to the compressibility.

\section{Exchange and correlation contributions to the compressibility} 
\label{sect:kappaxc}

\subsection{Formal electron-gas theory}
\label{sect:theory}

The compressibility $\kappa$ is defined by~\cite{Giuliani_and_Vignale}
\begin{equation}
\frac{1}{\kappa} = n^2 \frac{\partial \mu}{\partial n} =  \frac{n^2}{S} \frac{\partial^2 E}{\partial n^2}~,
\end{equation}
where $\mu = \partial E/\partial N$ is the chemical potential of the interacting system, $E$ is the total ground-state energy, and $n$ is the total (electron) density~\cite{phsymmetry}. 

Using the Hellman-Feynman coupling-constant-integration theorem~\cite{Giuliani_and_Vignale} 
and the specific form of ${\hat {\cal H}}_{\rm int}$ given above in Eq.~(\ref{eq:intenergy}) we find that the interaction contribution to the ground-state energy of BLG 
is given by
\begin{equation}\label{eq:energy}
E_{\rm int}= \frac{N}{2}\sum_{\ell=\pm}\int_0^{1}d\lambda
\int \frac{d^2 {\bm q}}{(2\pi)^2}~V_\ell(q) \left[S^{(\lambda)}_{\ell}(q)-1\right]~,
\end{equation}
where $S^{(\lambda)}_{\pm}(q)$ are the even and odd parity electron static structure factors at coupling constant $\lambda$. 
Appealing to the fluctuation-dissipation theorem we find that 
\begin{equation}\label{eq:FDT}
S^{(\lambda)}_{\pm}(q) = -\frac{1}{\pi n}\int_0^{+\infty} d\Omega~\chi^{(\lambda)}_{\pm}(q,i\Omega)~.
\end{equation}
This form of the fluctuation-dissipation theorem takes advantage of the smooth behavior of the linear-response functions 
$\chi^{(\lambda)}_{\pm}(q,i\Omega)$ along the imaginary axis, which simplifies the task of 
performing accurate numerical wavevector and frequency integrals.  (Along this axis one does not have to worry about the the 
collective plasmon poles and subtle particle-hole continuum band-edge features 
which are present along the real-frequency axis.~\cite{borghi_prb_2009}) 

Substituting Eq.~(\ref{eq:FDT}) in Eq.~(\ref{eq:energy}) we obtain an expression for the total ground-state energy of the interacting system:
\begin{eqnarray}\label{eq:energy-explicit}
E &=& E_0+\frac{N}{2} \sum_{\ell= \pm} \int_0^{1}d\lambda
\int \frac{d^2 {\bm q}}{(2\pi)^2}~V_\ell(q) \nonumber \\
&\times&\left[-\frac{1}{\pi n}\int_0^{+\infty} d\Omega~\chi^{(\lambda)}_{\ell}(q,i\Omega)-1\right]~,
\end{eqnarray}
$E_0$ being the trivial noninteracting kinetic energy. 
Following the conventional procedures of electron-gas theory~\cite{Giuliani_and_Vignale,barlas_prl_2007}, we separate out the contribution that is first order in $e^2$ ({\it i.e.} the ``exchange" energy) 
by writing $E = E_0+E_{\rm x} + E_{\rm c}$. The exchange energy per electron, $\varepsilon_{\rm x}=E_{\rm x}/N$, is given by
\begin{eqnarray}\label{eq:exchange}
\varepsilon_{\rm x} &=& \frac{1}{2}\sum_{\ell=\pm}
\int \frac{d^2 {\bm q}}{(2\pi)^2}~V_\ell(q) \nonumber\\
&\times&\left[-\frac{1}{\pi n}\int_0^{+\infty} d\Omega~\chi^{(0)}_{\ell}(q,i\Omega)-1\right]~,
\end{eqnarray}
where $\chi^{(0)}_{\pm}(q,i\Omega)$ are the response functions of the noninteracting system, which have been calculated in Ref.~\onlinecite{borghi_prb_2009}. The correlation energy (per electron), which by definition is the sum of all the terms of higher order in $e^2$, is given by
\begin{equation}\label{eq:correlation}
\varepsilon_{\rm c} = -\frac{1}{2 \pi n} \sum_{\ell=\pm}\int_0^{1}d\lambda
\int \frac{d^2 {\bm q}}{(2\pi)^2}\int_0^{+\infty}d\Omega~{\cal F}_\ell(q,i\Omega)~,
\end{equation}
where ${\cal F}_\ell(q,i\Omega) \equiv V_\ell(q) \Delta \chi^{(\lambda)}_{\ell}(q,i\Omega)$ and $\Delta \chi^{(\lambda)}_{\ell}(q,i\Omega) \equiv \chi^{(\lambda)}_{\ell}(q,i\Omega)-\chi^{(0)}_{\ell}(q,i\Omega)$. Neglecting correlations ({\it i.e.} treating interactions to first order in $e^2$) one obtains the HF result for the ground-state energy.~\cite{kusminskiy_prl_2008}

In this article we treat correlations within the random phase approximation (RPA)~\cite{Giuliani_and_Vignale} in which the response functions of the interacting system at coupling constant $\lambda$ are given by
\begin{equation}\label{eq:RPA}
\chi^{(\lambda)}_{\ell}(q,\omega) = \frac{\chi^{(0)}_{\ell}(q,\omega)}{1- \lambda V_\ell(q)\chi^{(0)}_{\ell}(q,\omega)}~.
\end{equation}
After inserting Eq.~(\ref{eq:RPA}) in Eq.~(\ref{eq:correlation}) one can observe that the integration 
over the coupling constant $\lambda$ can be performed analytically with the result
\begin{eqnarray}\label{eq:correlation-final-integrated}
\varepsilon^{\rm RPA}_{\rm c}&=&\frac{1}{2\pi n}\sum_{\ell = \pm 1}
\int \frac{d^2 {\bm q}}{(2\pi)^2}\int_0^{+\infty} d\Omega~\left\{V_\ell(q)\chi^{(0)}_{\ell}(q,i\Omega)\right. \nonumber \\
&+& \left. \ln\left[1-V_\ell(q)\chi^{(0)}_{\ell}(q,i\Omega)\right]\right\}~.
\end{eqnarray}

Since the hyperbolic bands of BLG asymptotically become linear in momentum (and thus identical to those of SLG), the integrals over frequency $\Omega$ in Eqs.~(\ref{eq:exchange}) and~(\ref{eq:correlation}) diverge. We thus proceed as in the single-layer case~\cite{barlas_prl_2007} and regularize the frequency integrals by subtracting from $\varepsilon_{\rm x}$ and $\varepsilon^{\rm RPA}_{\rm c}$ the {\it infinite} (and physically irrelevant) continuum modle exchange and RPA correlation energies of the {\it undoped} system. In this way we introduce a regularized exchange energy:
\begin{equation}\label{eq:exchange-regularized}
\delta \varepsilon_{\rm x} = -\frac{1}{2\pi n}\sum_{\ell = \pm 1}
\int \frac{d^2 {\bm q}}{(2\pi)^2}~V_\ell(q)\int_0^{+\infty} d\Omega~\delta \chi^{(0)}_{\ell}(q,i\Omega)~,
\end{equation}
where we have introduced the regularized response functions, $\delta \chi^{(0)}_{\ell}(q,i\Omega) \equiv \chi^{(0)}_{\ell}(q,i\Omega) - \chi^{(0{\rm u})}_{\ell}(q,i\Omega)$, $\chi^{(0{\rm u})}_{\ell}(q,i\Omega)$ being the noninteracting response functions of the undoped system~\cite{borghi_prb_2009}. The regularized correlation energy is given by
\begin{eqnarray}\label{eq:correlation-final-integrated-regularized}
\delta \varepsilon^{\rm RPA}_{\rm c}&=&\frac{1}{2\pi n}\sum_{\ell = \pm}
\int \frac{d^2 {\bm q}}{(2\pi)^2}\int_0^{+\infty} d\Omega~\Bigg\{V_\ell(q)\delta \chi^{(0)}_{\ell}(q,i\Omega)\nonumber\\
&+&
\ln\left[\frac{1-V_\ell(q)\chi^{(0)}_{\ell}(q,i\Omega)}{1-V_\ell(q)\chi^{(0{\rm u})}_{\ell}(q,i\Omega)}\right]\Bigg\}~.
\end{eqnarray}
This expression can be used to evaluate changes in interaction energy with carrier density at densities that are small compared to 
inverse unit cell area values, and is therefore reliable in the density range of relevance to gated and doped 
BLG electronic systems.  
Eqs.~(\ref{eq:exchange-regularized}) and~(\ref{eq:correlation-final-integrated-regularized}) are the most important results of this work. Together with the results for the doped $\chi^{(0)}_{\ell}(q,i\Omega)$ and undoped $\chi^{(0{\rm u})}_{\ell}(q,i\Omega)$ dynamical response functions presented in Ref.~\onlinecite{borghi_prb_2009} they allow us to accurately evaluate the ground-state energy (per electron) of BLG and thus the compressibility. 

\begin{figure}[h!]
\begin{center}
\begin{tabular}{c}
\includegraphics[width=9cm]{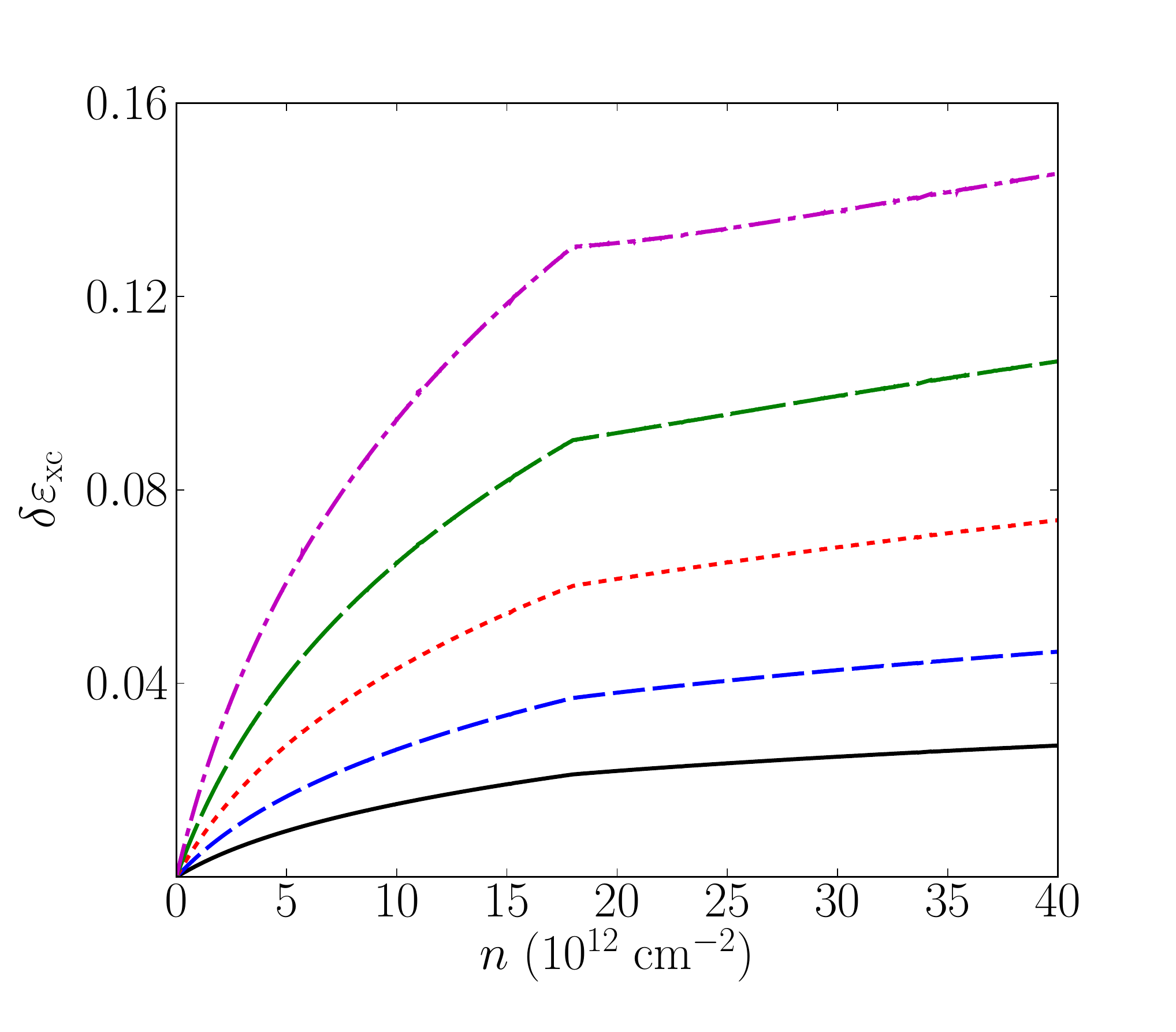} \\
\includegraphics[width=9cm]{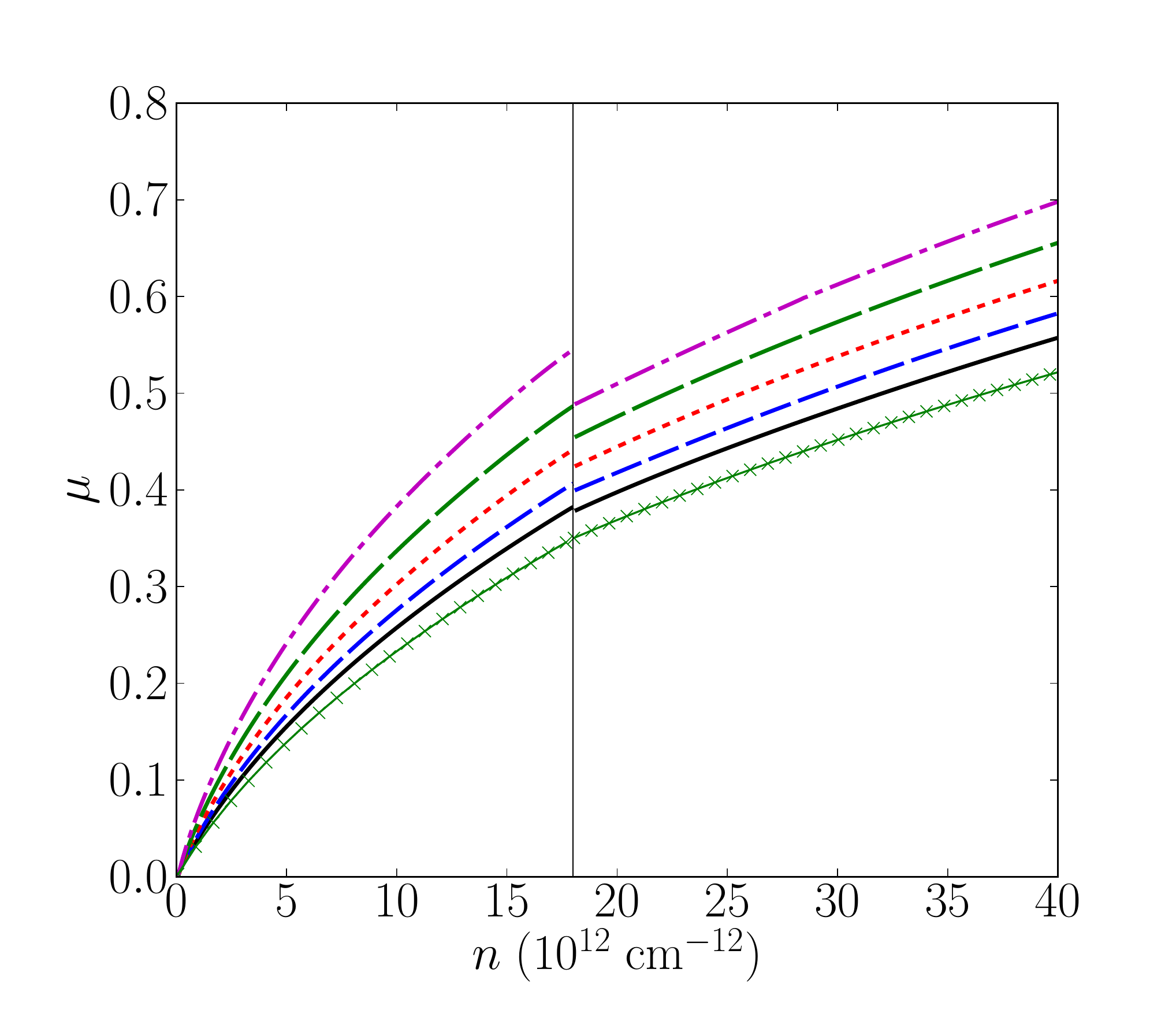}
\end{tabular}
\caption{(Color online) Top panel: interaction energy per electron (in ${\rm eV}$) 
as a function of doping (in units of $10^{12}~{\rm cm}^{-2}$) for different values of graphene's fine-structure constant 
$\alpha_{\rm ee}$. The values of $\alpha_{\rm ee}$ displayed are (from bottom to top) 
$\alpha_{\rm ee}=0.125$ (solid line), $0.25$ (short-dashed line), $0.5$ (dotted line), $1$ (long-dashed line), $2.2$ (dash-dotted line). Note the cusp for $n \approx 18 \times 10^{12}~{\rm cm}^{-2}$, which is more prominent for large values of $\alpha_{\rm ee}$, the value of doping at which the high-energy split off band $\varepsilon_1(k)$ is first occupied. Bottom panel: 
the chemical potential $\mu$ (in ${\rm eV}$) as a function of doping for different values of 
$\alpha_{\rm ee}$. Color-coding and labeling is the same as in top panel. Crosses label $\mu$ for $\alpha_{\rm ee} = 0$ (noninteracting bilayer graphene).\label{fig:one}}
\end{center}
\end{figure}

After the regularization procedure described above, the integrals over $\Omega$ in Eqs.~(\ref{eq:exchange-regularized}) and~(\ref{eq:correlation-final-integrated-regularized}) are finite but the ones over $q$ diverge. These divergences must be regularized by introducing an ultraviolet cut-off $q_{\rm max} \equiv \varepsilon_{\rm max}/v$. Below we choose $\varepsilon_{\rm F}/v$ as the 
unit of momentum, where $\varepsilon_{\rm F}$ is the Fermi energy:
\begin{equation}
\varepsilon_{\rm F} = \left\{
\begin{array}{l}
{\displaystyle v \sqrt{\frac{\pi}{2} n},~{\rm if}~\varepsilon_1(k)~{\rm is~occupied}}\\
{\displaystyle -\frac{t_\perp}{2}+\sqrt{\frac{t^2_\perp}{4} + v^2 \pi n },~{\rm if}~\varepsilon_1(k)~{\rm is~empty}}
\end{array}
\right.~.
\end{equation}
Thus the integrals over dimensionless wave vectors must be calculated up to a maximum value 
$\Lambda \equiv \varepsilon_{\rm max}/\varepsilon_{\rm F}$, corresponding to the 
highest energies at which the continuum model applies. We set $\varepsilon_{\rm max}$ to
\begin{equation}
\varepsilon_{\rm max} \equiv \sqrt{\frac{2\pi v^2}{{\cal A}_0}} = v \sqrt{\frac{\pi}{2} n_{\rm max}} \approx 7.2~{\rm eV}~,
\end{equation}
where ${\cal A}_0 = 3\sqrt{3} a^2_0/2$ is the SLG unit-cell area, $a_0 \approx 1.42$~\AA~being the carbon-carbon distance, and $n_{\rm max} = 4/{\cal A}_0$ is the number of $\pi$-electrons per unit area in the neutral system. The energies we evaluate have a weak logarithmic dependence on $\varepsilon_{\rm max}$ that has little importance for the conclusions we draw below.  

%
\begin{figure}[h!]
\begin{center}
\includegraphics[width=9cm]{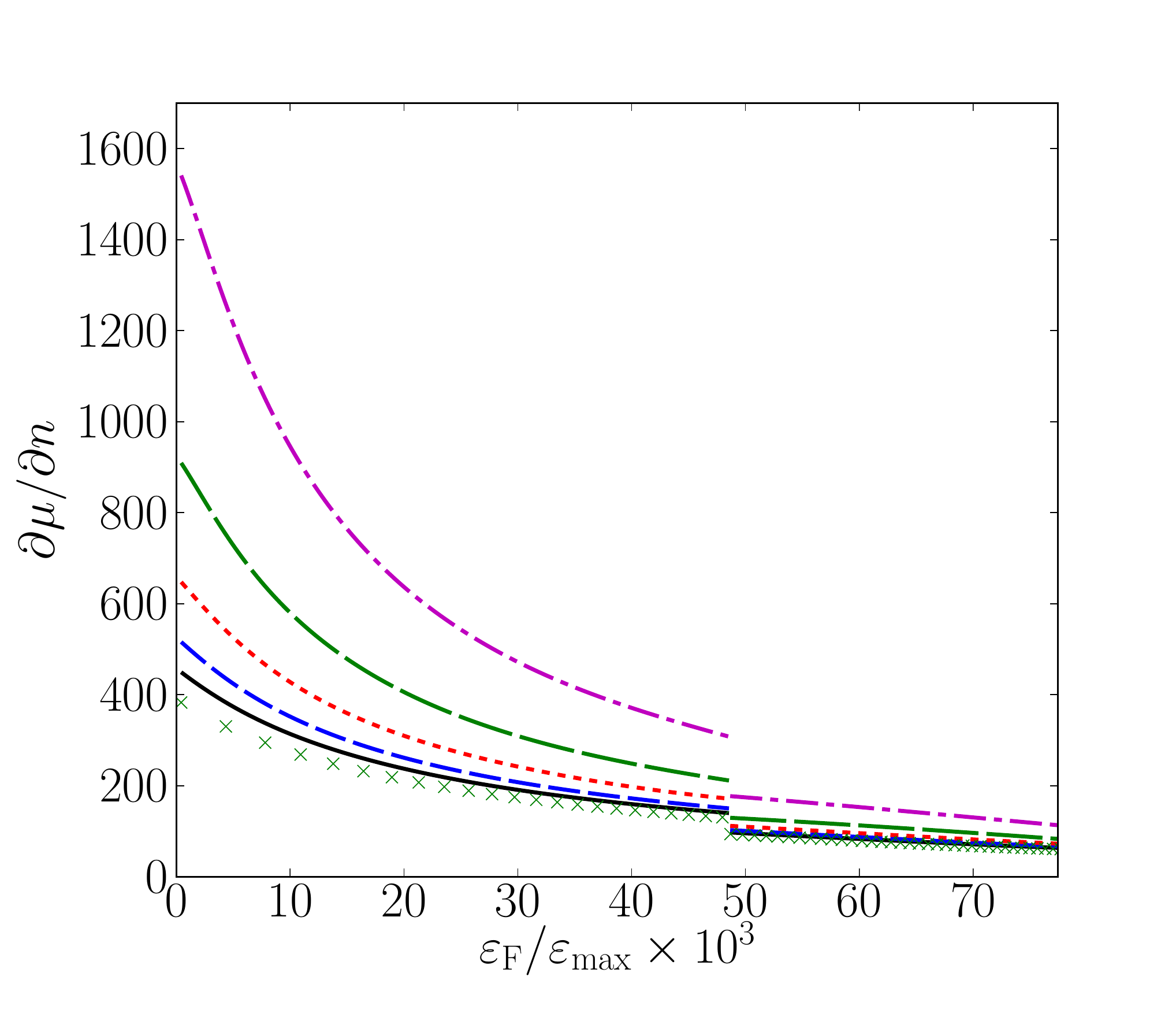}
\caption{(Color online) The Hartree-Fock (exchange-only) inverse 
thermodynamic density-of-states $\partial \mu/\partial n$ (in units of ${\rm eV} \times {\rm \AA}^2$) as a function 
of the Fermi energy $\varepsilon_{\rm F}$ (in units of $\varepsilon_{\rm max} \times 10^{-3}$) for different values of $\alpha_{\rm ee}$. 
The Fermi energy $\varepsilon_{\rm F}$ ranges from $\varepsilon_{\rm F} = 7\times 10^{-4}~{\rm eV}$ (corresponding to a doping $n\approx 2\times 10^{10}~{\rm cm}^{-2}$) to $\varepsilon_{\rm F} = 0.53~{\rm eV}$ (corresponding to a doping $n\approx 4.0 \times 10^{13}~{\rm cm}^{-2}$). Color-coding and labeling is the same as in Fig.~\ref{fig:one}. Crosses label $\partial \mu/\partial n$ for $\alpha_{\rm ee} = 0$ (noninteracting bilayer graphene). A negative $\delta$-function contribution to $\partial \mu/\partial n$ at $n = n_1$ 
for $\alpha_{\rm ee} \neq 0$ has been omitted.\label{fig:two}}
\end{center}
\end{figure}
%

%
\subsection{Numerical results and discussion}
\label{sect:mainresults}

We now turn to our main numerical results. The ground-state properties of BLG 
are completely determined by the total density $n$, by the interlayer distance $d$, which we have taken to be $d=3.35$~\AA, by the inter-layer hopping $t_\perp$, which we have taken to be $0.35~{\rm eV}$, and by the fine-structure constant (restoring $\hbar$ for a moment) 
$\alpha_{\rm ee} =e^2/(\hbar v \epsilon)$.

\begin{figure}[h!]
\begin{center}
\begin{tabular}{c}
\includegraphics[width=9cm]{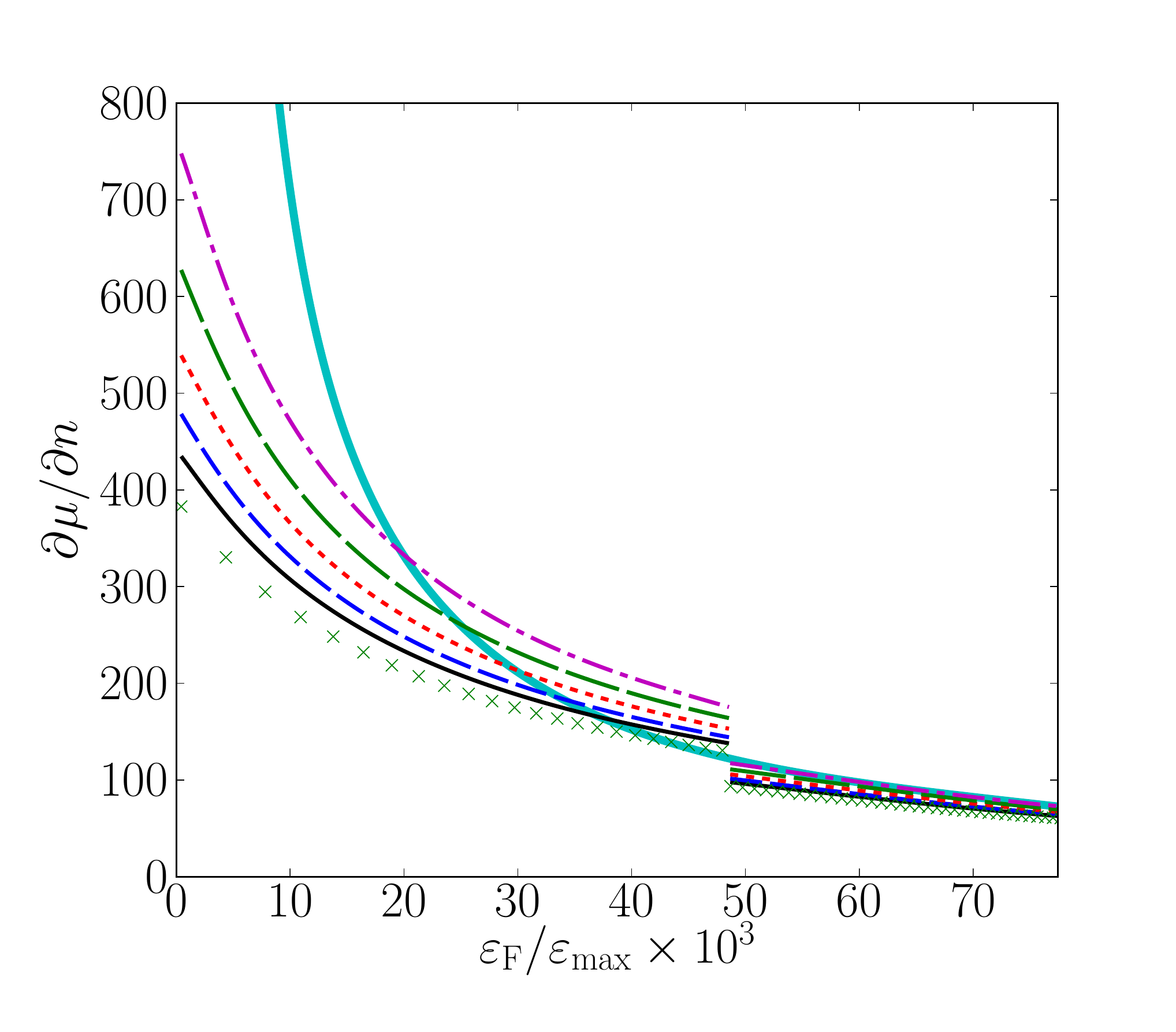}\\
\includegraphics[width=9cm]{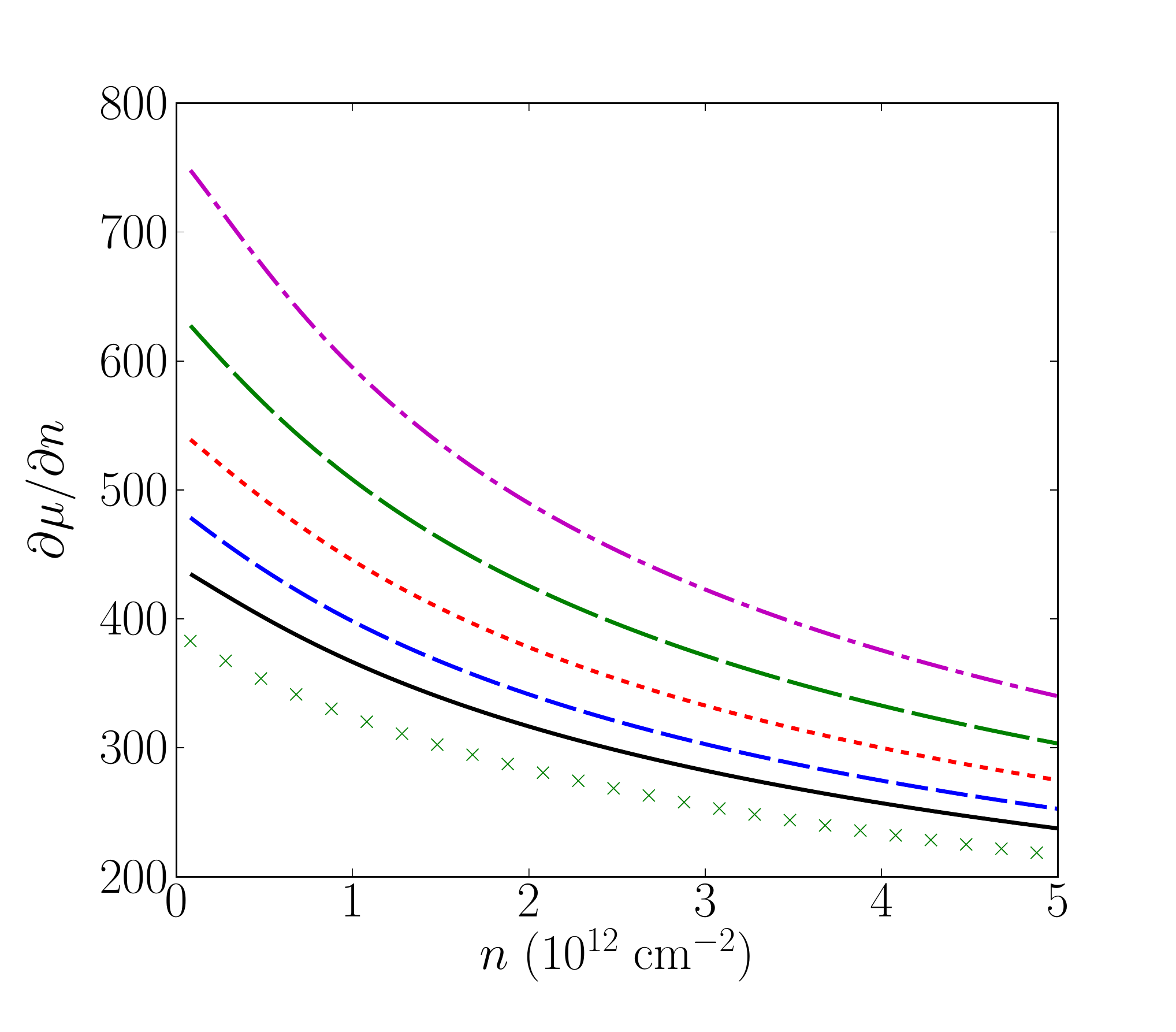}
\end{tabular}
\caption{(Color online) Top panel: same as in Fig.~\ref{fig:two} but with the inclusion of RPA correlations. The thick solid (cyan) line labels 
the inverse thermodynamic density-of-states of suspended ($\alpha_{\rm ee}=2.2$) single-layer graphene. The values of $\varepsilon_{\rm F}$ range from $\varepsilon_{\rm F}=7\times 10^{-4}~{\rm eV}$ to $\varepsilon_{\rm F}=0.53~{\rm eV}$ ($n\approx 4.0 \times 10^{13}~{\rm cm}^{-2}$). Note that the compressibility of bilayer graphene remains finite for $n \to 0$, while the one of single-layer graphene diverges. A negative $\delta$-function contribution to $\partial \mu/\partial n$ at $n = n_1$ for $\alpha_{\rm ee} \neq 0$ has been omitted.
Bottom panel: a zoom of the upper panel for low densities. Note that the horizontal axis in the bottom panel represents total density $n$ in units of $10^{12}~{\rm cm}^{-2}$. \label{fig:three}}
\end{center}
\end{figure}

\begin{figure}[t]
\begin{center}
\includegraphics[width=9cm]{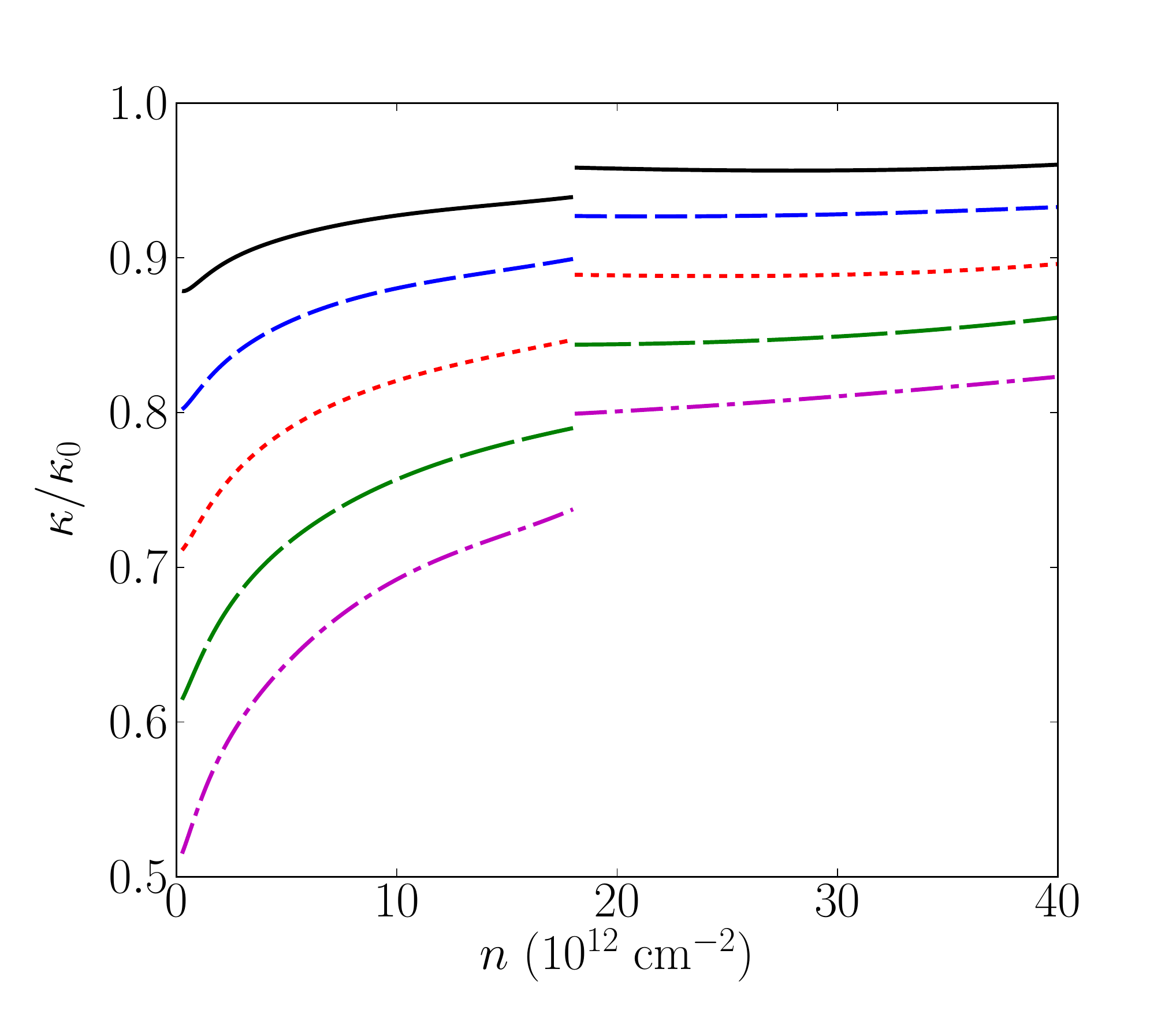}
\caption{(Color online) The ratio between the interacting-system compressibility ($\kappa$) and the noninteracting one ($\kappa_0$), 
as a function of doping (in units of $10^{-12}~{\rm cm}^{-2}$). Color coding and labeling are the same as in Figs.~\ref{fig:two}-\ref{fig:three}. Note that the ratio $\kappa/\kappa_0$ is smaller than unity, the more so the stronger electron-electron interactions are.\label{fig:four}}
\end{center}
\end{figure}

In the top panel of Fig.~\ref{fig:one} we present the exchange-correlation energy $\delta \varepsilon_{\rm xc} \equiv \delta \varepsilon_{\rm x} + \delta \varepsilon^{\rm RPA}_{\rm c}$ as a function of $n$ and $\alpha_{\rm ee}$.  For our choice of energy zero, $\delta \varepsilon_{\rm x}$ is positive and $\delta \varepsilon_{\rm c}$ is negative.  The two contributions to the interaction energy tend to cancel strongly, with a slightly positive total, suggesting that correlation will be as important as exchange in determining physical properties.   We see that $\delta \varepsilon_{\rm xc}$ has a cusp at every value of $\alpha_{\rm ee}$ for $ n= n_1 \equiv 2(t_\perp/v)^2/\pi \approx 18 \times 10^{12}~{\rm cm}^{-2}$, the value of doping at which $\varepsilon_{\rm F} = t_\perp$.  It is at this value of $n$ that the high-energy split off band $\varepsilon_1(k)$ is first occupied.  As a consequence, the chemical potential $\mu = \partial [n (\delta \varepsilon_{\rm kin} + 
\delta \varepsilon_{\rm xc})]/\partial n$ has a jump at $n =n_1$ when doping is increased from values smaller than $n_1$ to values larger than $n_1$. Here $\delta \varepsilon_{\rm kin}$ is the kinetic energy (per electron) of the noninteracting system with the Dirac point chosen as energy zero. The chemical potential $\mu$ as a function of doping is illustrated in the bottom panel of Fig.~\ref{fig:one}. Note that the jump is {\it downward}. These type of jumps have been discussed earlier in the context of second-subband occupation in wide quantum wells (see for example Ref.~\onlinecite{jo_prb_1993} and references therein) and are potentially technologically interesting since they can in principle lead to bistability. We observe that the chemical potential jump implies a $\delta$-function singularity in the compressibility $\kappa$ at $n = n_1$: this feature will be omitted in the presentation of compressibility numerical results below, Figs.~(\ref{fig:two})-(\ref{fig:four}).

In Fig.~\ref{fig:two} we report HF theory results for the inverse 
thermodynamic density-of-states $\partial \mu/\partial n$ which has kinetic and exchange energy contributions:  
$\partial \mu/\partial n|_{\rm HF} \equiv \partial^2 [n(\delta \varepsilon_{\rm kin}+\delta \varepsilon_{\rm x})]/\partial n^2$.
The decrease in $\partial \mu/\partial n$ with density at $\alpha_{\rm ee} = 0$ is a consequence of the 
difference between hyperbolic and parabolic dispersion.  We see that $\partial \mu/\partial n$ is 
positive and enhanced by exchange interactions over the density range covered in this plot. 
The non-monotonic behavior predicted in Ref.~\onlinecite{kusminskiy_prl_2008} appears only at 
extremely low densities and, as we discuss below, does not survive RPA correlations.
This behavior contrasts with that of ordinary 2D electron gases in which 
HF theory predicts a negative compressibility below a critical density that is 
moderate, and only weakly influenced by correlations.  This qualitative behavior difference is a consequence of 
the relevance of exchange interactions with both conduction and valence bands, and of the sublattice pseudospin 
chirality of the bands.  The same mechanisms are also responsible for a thermodynamic density-of-states that is 
suppressed rather than enhanced in SLG~\cite{barlas_prl_2007,borghi_ssc_2009}.
Like an ordinary 2D electron gas, BLG has an intra-conduction-band exchange contribution to its chemical potential that is negative
and proportional to $n^{1/2}$; this energy is however approximately half as large in the BLG case because the 
wavefunctions are spread over more than one sublattice.  The tendency toward a negative compressibility is 
further countered in the BLG case by inter-band exchange, which yields a chemical potential 
contribution that is also proportional to $n^{1/2}$ in the low-density limit, but positive.  The end result is that the low-carrier 
density negative $n^{1/2}$ exchange contribution to the chemical potential, which would yield 
a negative compressibility, is approximately six times 
smaller (for the same background dielectric constant) in BLG than in an ordinary 2D electron gas.  When only 
exchange interactions are included, 
we find that the total compressibility calculated within the two-band model~\cite{mccann_prl_2006} becomes negative 
only at densities below a critical value given by (restoring $\hbar$ for a moment)
\begin{equation}\label{eq:critdens}
n_{\rm c} = \frac{1}{\pi} \left(\frac{\beta \alpha_{\rm ee} t_\perp}{2 \hbar v}\right)^2 \approx \alpha^2_{\rm ee} (1.1\times 10^{10})~{\rm cm}^{-2}~,
\end{equation}
in agreement with the numerical results in Fig.~2 of Ref.~\onlinecite{kusminskiy_prl_2008}. In Eq.~(\ref{eq:critdens})
we have introduced
\begin{equation}\label{eq:betafactor}
\beta =  \frac{1}{\pi} - \frac{3}{16}\int_{1}^{\infty}dx \frac{1}{x^2}~_2F_1(5/2,1/2,3,1/x^2) = \frac{2}{9\pi}~,
\end{equation}
$_2F_1(a,b,c,x)$ being the usual hypergeometric function. For the sake of comparison, note that within HF the critical density at which the compressibility of a standard 2D electron gas changes sign is $n^{({\rm 2DEG})}_{\rm c} = 2/(\pi^3 a^2_{\rm B})$, where $a_{\rm B}$ is the material Bohr radius. In GaAs, for example, $a_{\rm B} \approx 100$~\AA~and thus $n^{({\rm 2DEG})}_{\rm c} \approx 6.5 \times 10^{10}~{\rm cm}^{-2}$. In the graphene case we find numerically that in the random phase approximation the low-density 
negative compressibility does not survive correlations.  This compressibility anomaly is in any event
likely to be preempted by BLG's  low-density ferroelectric instability~\cite{min_prb_2008,zhang_prb_2010,nandkishore_prl_2010}, which is driven by physics beyond that captured by the RPA as discussed above. The issue of a possible negative compressibility in BLG is discussed further below.

In Fig.~\ref{fig:three} we report on results for the inverse thermodynamic density-of-states, $\partial \mu/\partial n$,
calculated including both exchange and RPA correlations corrections: $\partial \mu/\partial n\equiv \partial^2 [n(\delta \varepsilon_{\rm kin}+\delta \varepsilon_{\rm xc})]/\partial n^2$. Qualitatively, the results in Fig.~\ref{fig:three} look 
rather similar to the HF ones in Fig.~\ref{fig:two}.

However, we clearly see that for dopings below $n_1 \approx 18 \times 10^{12}~{\rm cm}^{-2}$ 
correlation effects are quantitatively very important. For instance, percentage values of the ratio
\begin{equation}
r(\alpha_{\rm ee}) = \lim_{n \to 0} \frac{\partial \mu/\partial n}{(\partial\mu/\partial n)|_{\rm HF}}~,
\end{equation}
(between the data in Fig.~\ref{fig:three} and  the HF data in Fig.~\ref{fig:two}) are of the order of $\approx 20$\% for $\alpha_{\rm ee} = 0.5$ and larger than $100$\% for a suspended bilayer ($\alpha_{\rm ee} = 2.2$). 

For the sake of comparison in Fig.~\ref{fig:three} we have also plotted $\partial \mu/\partial n$ for suspended ($\alpha_{\rm ee} = 2.2$) SLG. As expected, the difference between DLG and SLG compressibilities is very small at high doping, especially so when all four bands $\varepsilon_i(k)$ are occupied. At low densities, however, the results are very different since in this regime the BLG spectrum approaches a parabolic form $k^2/(2m)$, with $m = t_{\perp}/(2v^2)$, strongly deviating from the SLG linear dispersion. In particular, note that $\partial \mu/\partial n$ diverges for $n \to 0$ in the SLG case, while it approaches a finite value for BLG. This striking difference stems from the behavior of the BLG quasiparticle effective mass $m^\star$, which remains finite when doping approaches zero~\cite{borghi_ssc_2009}.

In Fig.~\ref{fig:four} we plot the ratio $\kappa/\kappa_0$, $\kappa_0= [n^2 \partial^2 (n \delta \varepsilon_{\rm kin})/\partial n^2]^{-1}$ being the compressibility of the noninteracting system. We clearly see from this plot that the main effect of electron-electron interactions is to suppress $\kappa$. This can be easily understood within the Landau theory of normal Fermi liquids. Indeed $\kappa/\kappa_0$ is largely controlled by and proportional to $m^\star/m$ and, as demonstrated in Ref.~\onlinecite{borghi_ssc_2009}, the role of interactions is to suppress $m^\star$ with respect to the bare value, {\it i.e.} $m^\star/m <1$. As explained in Ref.~\onlinecite{barlas_prl_2007} for the case of SLG, the suppression of the mass (or enhancement of the quasiparticle velocity) stems from the chiral nature of the low-energy spectrum.
%
%

\begin{figure}[t]
\begin{center}
\includegraphics[width=9cm]{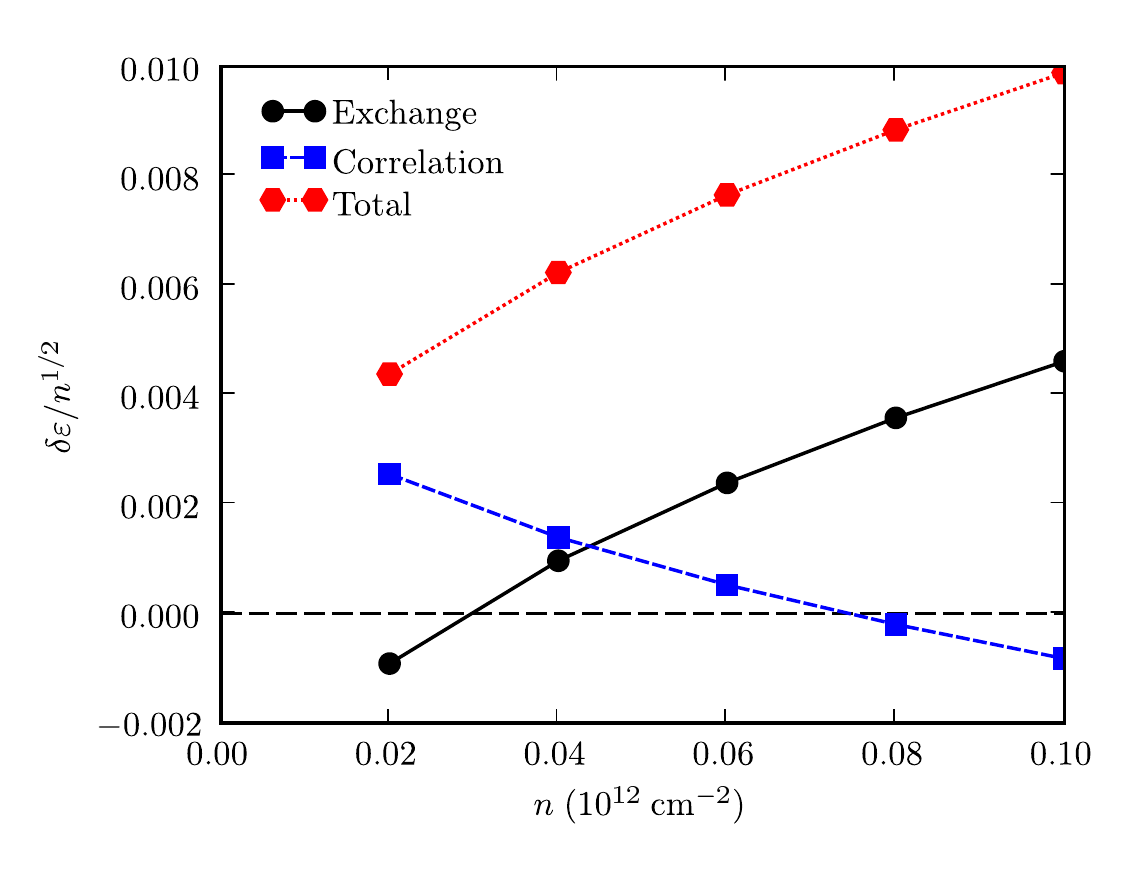}
\caption{(Color online) The various contributions to the ground-state energy (per particle and divided by $n^{1/2}$) as functions of the carrier density $n$ in units of $10^{12}~{\rm cm}^{-2}$. These results refer to $\alpha_{\rm ee} = 1$.\label{fig:five}}
\end{center}
\end{figure}

We now turn to a discussion of the compressibility in the extreme low-density limit, illustrated in Fig.~\ref{fig:five}.  As discussed above and first explained by Kusminskiy {\it et al.}~\cite{kusminskiy_prl_2008}, the compressibility becomes negative at extremely low densities in the HF approximation because of a small net contribution to the chemical potential that is negative and proportional to $n^{1/2}$.  This contribution to the compressibility is reminiscent of the larger but related contribution to the chemical potential which appears in an ordinary 2D electron gas.  Because the relative strength of interactions and band-energies in that case can be absorbed in a length scale change, the $n^{1/2}$ exchange energy can be viewed as the leading order term in an expansion of energy in powers of $e^2/k_{\rm F} \sim e^2 n^{-1/2}$.  The leading order correlation contribution to the chemical potential is therefore proportional to $n^{0}$ (up to logarithmic factors) and does not appear in the compressibility.  This simple scaling property does not apply to BLG, because the inter-layer hopping and in-plane hopping terms in the continuum-limit Hamiltonian do not scale in the same way with density.  The chemical potential and energy per particle have to be expanded separately in terms of powers of $n^{1/2}$ and the interaction scale $\alpha_{\rm ee}$.  As illustrated in Fig.~\ref{fig:five} we find numerically that both the exchange and correlation contributions to the chemical potential change sign at very low carrier densities, in such a way that the total chemical potential is a monotonically increasing function of energy.  In fact we find that the ratio $\kappa/\kappa_0$ is smallest at low density: in other words, as it is also clear from Figs.~\ref{fig:two}-\ref{fig:three}, the enhancement of $\partial \mu/\partial n$ relative to the noninteracting system results becomes larger for lower carrier densities. 

The competing role of exchange and correlations here is similar to their role in the spin-susceptibility.  The exchange energy favors spin-polarization by lowering the chemical potential of each spin when its occupation increases.  When correlations are included, the energy depends more on the total density and less on its partitioning into spin components.  Similarly here the exchange energy favors occupation of the higher subband at low-densities because of the large Coulomb interaction matrix elements when the Fermi radius is small.  When correlations - which are less sensitive to interactions between quasiparticles at the Fermi energy - are restored, the sensitivity to band index is reduced and the overall trend in the dependence of chemical potential on density is restored.

%
%

%
\section{Summary and discussion}
\label{sect:conclusions}

In summary, we have calculated the compressibility of crystalline bilayer graphene beyond the Hartree-Fock approximation by treating correlation effects at the random-phase-approximation level. 
We have shown that electron-electron interactions suppress the compressibility quite substantially with correlation effects playing an important quantitative role. The reduction of compressibility stands in stark contrast to the large compressibility enhancements that occur in regular two-dimensional electron gas systems, even though the two systems share the same parabolic dispersions.  The source of the qualitatively different behavior is the importance in bilayer graphene of exchange interactions between carriers in the conduction band and the full negative energy Dirac sea.  The suppression of the compressibility has the same origin as the enhancement of quasiparticle effective mass.
Both phenomena ultimately originate from the chiral nature of the low-energy spectrum.

The present results demonstrate that correlations play an essential role in quantitative studies of 
interaction effects in bilayer graphene.  Previous work~\cite{min_prb_2008,zhang_prb_2010,nandkishore_prl_2010} 
has suggested that neutral bilayers might become unstable to spontaneous layer polarization when disorder is weak.  
The role of long-range Coulomb interactions in the physics of this instability could be addressed by extending the calculations
described here to the case in which there is an electric potential difference between layers.
One complication associated with this elaboration is that inversion symmetry is explicitly 
broken so that correlations between even and odd parity density fluctuations are non-zero.  The more complicated form for the 
subband spinors of the band eigenstates also makes the task of finding quasi-analytic 
results for the noninteracting system polarization (Lindhard) functions challenging. 
In the present calculation the semi-analytic results we have used for the Lindhard function~\cite{borghi_prb_2009} 
are extremely helpful and allow the 
wavevector and frequency integrals to be evaluated numerically with confidence and precision, in spite of the numerical subtleties that lurk in the integrands.  It is difficult to obtain accurate results when the Lindhard function is evaluated by brute-force numerics.   
Although this lies outside the scope of the present paper, we anticipate that 
correlation effects will lower predictions for the amount of charge transferred 
between the layers.  

After this work was complete we learned of three recent experimental studies~\cite{Henriksen,Young,Yacoby} which measure 
the compressibility of BLG. All three groups find that, in the balanced limit, 
$\partial \mu/\partial n$ has a peak near zero carrier density and then decreases monotonically 
with increasing carrier density; the change of sign at low densities which appears in an ordinary two-dimensional 
electron gas is absent in BLG, in agreement with our findings.  Our calculations demonstrate that these 
experimental results can be strongly influenced by interactions so that some caution must 
be exercised in fitting compressibility measurements to band-structure models.      

\acknowledgments
We thank Antonio Castro Neto, Jim Eisenstein, Erik Henriksen, Amir Yacoby, and Andrea Young for fruitful discussions and correspondence.
M.P. acknowledges partial support by the 2009/2010 CNR-CSIC scientific cooperation project.
A.H.M. acknowledges support by the Welch Foundation and by Department of Energy grant no. DE-FG03-02ER45958 
(Division of Materials, Sciences, and Engineering).

\appendix

\section{The unitary transformation ${\cal U}_{\bm k}$}
\label{app:unitaryU}

For the sake of completeness, in this Appendix we report the form of the unitary matrix ${\cal U}_{\bm k}$ which diagonalizes the kinetic matrix ${\cal T}({\bm k})$. The matrix ${\cal U}_{\bm k}$ can be written as
\begin{equation}\label{eq:Umatrix}
{\cal U}_{\bm k} = G(\phi_{\bm k}) A R(\theta_{\bm k})~,
\end{equation}
where the two angles $\phi_{\bm k}$ and $\theta_{\bm k}$ are defined by $\phi_{\bm k}=\arctan(k_y/k_x)$ and $\theta_{\bm k}=\arctan[\sqrt{\varepsilon_3(\bm k)/\varepsilon_1(\bm k)}]$. 

The $4 \times 4$ matrices $G(\phi_{\bm k})$, $A$, and $R(\theta_{\bm k})$ in Eq.~(\ref{eq:Umatrix}) are given by
\begin{equation}
G_{\alpha\beta}(\phi_{\bm k}) = \delta_{\alpha\beta} [\delta_{\alpha 1}+\delta_{\alpha 4} - e^{i\phi_{\bm k}} \delta_{\alpha 2} 
- e^{-i\phi_{\bm k}} \delta_{\alpha 3}]~,
\end{equation}
$\delta_{ij}$ being the usual Kronecker delta, 
\begin{equation}
A = \frac{1}{\sqrt{2}}(\gamma^5 + \gamma^5 \gamma^x)~,
\end{equation}
and, finally, $R(\theta_{\bm k})$ is the tensor product of two $2\times2$ rotations which act on the subspace spanned by the odd (inversion-antisymmetric) 
and even (inversion-symmetric) eigenstates of the kinetic Hamiltonian ${\hat {\cal T}}$, respectively. 
Recalling that the odd bands are those corresponding to the eigenvalues $\varepsilon_{1,2}(k)$, while the even ones are those labeled by the eigenvalues $\varepsilon_{3,4}(k)$, we find that $R(\theta_{\bm k}) = R_2(\theta_{\bm k})|_{1,2}\otimes R_2(\theta_{\bm k})|_{3,4}$ with
\begin{equation}
R_2(\theta_{\bm k})=\left(
\begin{array}{cc}
\cos(\theta_{\bm k}) &\sin(\theta_{\bm k})\\
-\sin(\theta_{\bm k}) & \cos(\theta_{\bm k})
\end{array}\right)~.
\end{equation}


\begin{thebibliography}{77}
\bibitem{notationfootnote}
	In this article we are using a slightly simplified notation with respect to that used in Ref.~\onlinecite{borghi_prb_2009}. 
	The total-density response function, here labeled by $\chi_+$, is identical to what there was labeled by $\chi_{\rho\rho}$.
	Similarly, the density-difference response function, here labeled by $\chi_-$, was there identified by the symbol $\chi_{\Upsilon\Upsilon}$.
\bibitem{phsymmetry}
	The model's particle-hole symmetry guarantees that electron-doped and hole-doped bilayers have identical properties.
\end{thebibliography}
\end{document}